\documentclass[aps,pre,reprint,groupedaddress,showpacs,twocolumn]{revtex4}
\usepackage {graphicx}
\usepackage{amsmath}

\begin{document}

\title{Phase diagram of truncated tetrahedral model}

\author{Roman \textsc{Krcmar}$^{1}$}
\author{Andrej \textsc{Gendiar}$^{1}$}
\email[]{andrej.gendiar@savba.sk}
\author{Tomotoshi \textsc{Nishino}$^{2}$}
\affiliation{$^1$Institute of Physics, Slovak Academy of Sciences, SK-845 11, Bratislava, Slovakia}
\affiliation{$^2$Department of Physics, Graduate School of Science, Kobe University, Kobe 657-8501, Japan}

\date{\today}

\begin{abstract}
Phase diagram of a discrete counterpart of the classical Heisenberg model, the truncated 
tetrahedral model, is analyzed on the square lattice, when the interaction is ferromagnetic. 
Each spin is represented by a unit vector that can point to one of the 12 vertices of the truncated 
tetrahedron, which is a continuous interpolation between the tetrahedron and the octahedron. 
Phase diagram of the model is determined by means of the statistical analogue of the 
entanglement entropy, which is numerically calculated by the corner transfer matrix 
renormalization group method. The obtained phase diagram consists of four different phases, 
which are separated by five transition lines. In the parameter region, where the octahedral 
anisotropy is dominant, a weak first-order phase transition is observed.
\end{abstract}

\maketitle

%%%%%
\section{Introduction}
%%%%%
Symmetry breaking is one of the fundamental concepts in the field theoretical analyses. 
Phase transitions in statistical models are known as the typical realizations of the symmetry 
breaking, where the feature of transitions is dependent on symmetries in local degrees 
of freedom. For example, the classical Heisenberg model has the $O( 3 )$ symmetry, 
and there is ferromagnetic-paramagnetic phase transition of the second order when the 
model is on the cubic lattice, and when the interaction is ferromagnetic. In this case the
transition temperature is of the order of the interaction energy divided by Boltzmann constant.

The $O(3)$ symmetry group has discrete subgroups, some of which correspond to polyhedral 
symmetries that correspond to Platonic polygons. Discrete counterpart of the classical Heisenberg 
model can be defined according to the polyhedral group symmetries. For example, a ferromagnetic 
30-state discrete vector spin model was introduced by Rapaport, for the purpose of simplifying 
Monte-Carlo simulation of the three-dimensional (3D) ferromagnetic Heisenberg 
model~\cite{Rapaport}. In this discretization, the middle points of the edges of the icosahedron, 
which has 12 vertices and 30 edges, are allowed to be the local spin degrees of freedom. 
It was shown that the calculated phase transition temperature coincides well with that of the 
3D Heisenberg model. Margaritis {\it et al.} considered a 12-state discrete vector model, which 
corresponds to the icosahedral symmetry, and also the 20-state one with the dodecahedral 
symmetry~\cite{Margaritis}. On the cubic lattice, it was shown that the 12-state model already 
well represent the phase transition of the 3D Heisenberg model. Thus, in three dimensions, the 
effect of such discretization introduced in the 12-, 20-, and 30-state vector models is irrelevant, as
long as the universality of the phase transition is concerned.

In two dimensions (2D), the situation is somewhat different. The continuous symmetry of the 
Heisenberg model prohibits the phase transition at finite temperature when the model is defined 
on 2D lattices~\cite{merminwagner}. Thus, if a discrete symmetry is introduced, it could be 
a relevant 
perturbation. It is instructive to remind that the $q$-state clock model, the discrete analogue of the 
classical XY model, shows a Bereziskii-Kosterlitz-Thouless (BKT) phase transition~\cite{B1, B2, KT}
when $q \ge 5$. In case of the ferromagnetic 
tetrahedral model on the square lattice, where only 4 states are allowed, there is a phase transition subject to the 4-state Potts universality class~\cite{wu}.
Nienhuis {\it et al.} showed that the cubic anisotropy is relevant to the $O(3)$ symmetry on 2D lattice, 
and a nontrivial phase diagram was reported for the ferromagnetic case~\cite{Nienhuis}.
Margaritis {\it et al.} confirmed the presence of the order-disorder phase transition in the discrete vector 
spin models with 12, 20, and 30 degrees of freedom on the square lattice and showed that the 
transition temperature is strongly dependent on the number of the local spin states~\cite{Margaritis}. 
Patrascioiu and Seiler performed a scaling analysis for the case of the icosahedral discretization and 
estimated the critical exponents assuming that the transition is of the second order~\cite{Patrascioiu}. 
A perturbative analysis of the critical behavior has been performed by Caracciolo {\it et al.}, and critical indices 
for the tetrahedral, cubic, and octahedral cases were estimated~\cite{Caracciolo1,Caracciolo2}.
Surungan revisited the icosahedral and the dodecahedral cases and obtained transition
temperature and critical exponents, that agree with the previous studies~\cite{surungan}.

A theoretical interest in the 2D polyhedral models is being focused on cases, where the discrete
symmetry group has subgroups. In such cases, successive transitions from a phase with a higher 
symmetry to another phase with a lower symmetry can be observed; the symmetry is only partially 
broken in the intermediate temperature region. Surungan {\it et al.} investigated a discrete counterpart 
of the Heisenberg model, the edge-cubic model, in which the local spins can point to one of the 12 vertices of the cuboctehadron~\cite{okabe}.  They detected two phase transitions; the first one in the low-temperature side 
of the 3-state Potts universality class, and the second one in the high-temperature side, 
which could be explained by the cubic symmetry. The edge-cubic 
model belongs to a variety of truncated Platonic --- Archimedean --- solid models, which can
also be regarded as 
discrete counterparts to the classical Heisenberg model. In this work we investigate another 
example of the truncated models, the truncated tetrahedral model (TTM), which is defined as a continuous interpolation 
between the tetrahedral and the octahedral models. The reason we have chosen this case of TTM is that the octahedral case with the 6 degrees of freedom is much less studied if compared to other symmetries, and we intend to analyze the stability of the critical behavior with respect to perturbations toward the tetrahedral symmetry.

This article is organized as follows. In the next section we introduce the TTM, 
and briefly discuss its property around the octahedral and the tetrahedral limits.
In Section~III, the phase diagram in the entire parameter region is determined by means
of the classical analogue of the entanglement entropy, which is calculated from the 
spectrum of the density matrix obtained by the Corner Transfer Matrix Renormalization
Group (CTMRG) method~\cite{ctmrg1,ctmrg2}.  We then classify the nature of the phase
transition lines.  The obtained results are summarized in the last section.

%%%%%
\section{Truncated Tetrahedral Model}
%%%%%

%
\begin{figure}
\includegraphics[width=0.40\textwidth,clip]{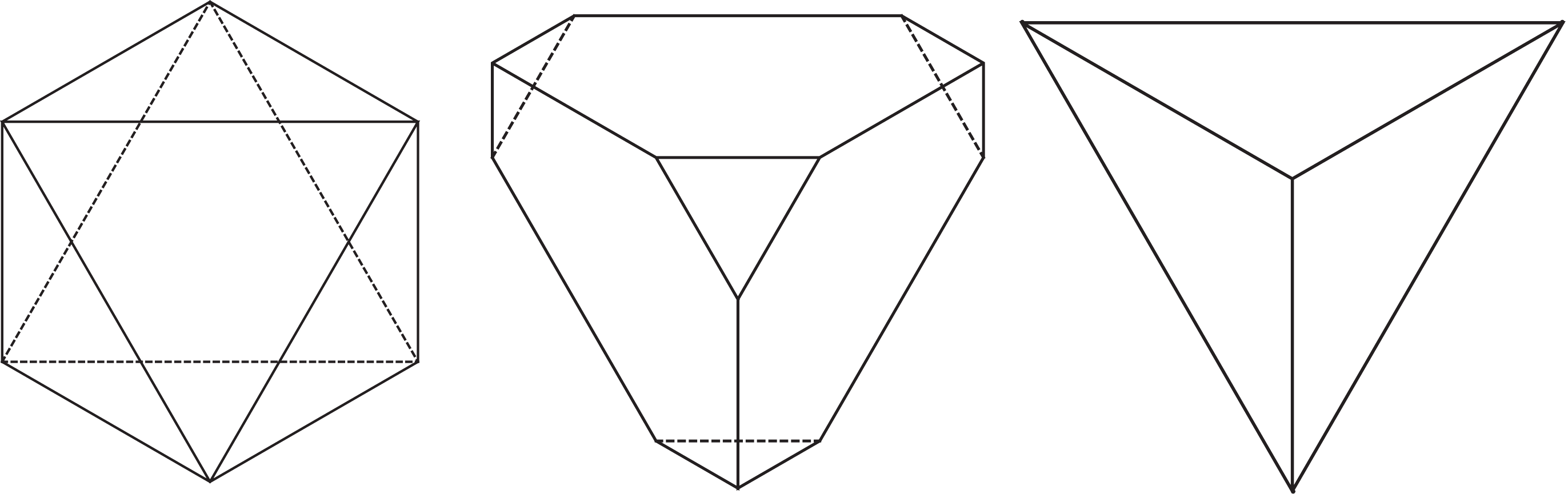}
\caption{
The truncated tetrahedron (shown in the middle, parameterized by $t = 0.5$) 
is depicted as the interpolation between the octahedron
(on the left for $t = 0$) and the tetrahedron (on the right for $t = 1$). 
}
\label{fig1}
\end{figure}

We consider a finite-directional counterpart of the classical Heisenberg model on
the 2D square lattice. Local spin on each lattice site is represented by a unit vector 
that can point to one of the 12 vertices of the truncated tetrahedron shown in
Fig.~\ref{fig1}, whose shape is 
determined by a parameter $t$ from octahedral limit $t=0$ to the tetrahedral one $t=1$. 
We represent the spin variable on the site that is specified by indices $i$ and $j$ 
by means of the unit vector $\mathbf{S}_{i,j}^{~}( t )$ given by
\begin{equation}
\mathbf{S}_{i,j}^{~}( t ) = 
\sqrt{ \frac{1 + 2 t^2_{~}}{2} } \,\, \mathbf{s}^{[ k ]}_{~}( t ) \, ,
\end{equation}
where the components of unnormalized vector $\mathbf{s}^{[ k ]}_{~}( t )$ for $k = 1 
\sim 12$ are listed in Table~\ref{tab1}. 
We assume that ferromagnetic coupling $J > 0$ is present between nearest-neighbor 
sites, and that the interaction is represented in the form of inner product. 
Under these settings, the Hamiltonian of the TTM is written as
\begin{equation}
H_t^{~} = - J \sum_{i, \, j}^{~}
 \biggl[
    \mathbf{S}_{i,j}^{~}( t ) \cdot \mathbf{S}_{i+1,j}^{~}( t )
  + \mathbf{S}_{i,j}^{~}( t ) \cdot \mathbf{S}_{i,j+1}^{~}( t )
 \biggr] \, .
\end{equation}
\begin{table}
\caption{The directions of the 12 vertices of the truncated tetrahedron 
represented by means of the unnormalized vector $\mathbf{s}^{[ k ]}_{~}( t )$.}
\begin{center}
\renewcommand{\arraystretch}{1.5}
\begin{tabular}{|c|rrr||c|rrr|}
\hline
$k$ & \multicolumn{3}{ c ||} {$\mathbf{s}^{[k]}_{~}(t)$} &
$k$ & \multicolumn{3}{ c | } {$\mathbf{s}^{[k]}_{~}(t)$} \\
\hline
$  1 $ & $\Big(\hfill            t,$ & $ 0,            $ & $-\frac{1}{\sqrt{2}}\Big)$ &
$  2 $ & $\Big(\hfill           -t,$ & $ 0,            $ & $-\frac{1}{\sqrt{2}}\Big)$ \\[0.1cm]

$  3 $ & $\Big(\hfill\frac{1-t}{2},$ & $ \frac{1+t}{2},$ & $ \frac{t}{\sqrt{2}}\Big)$ &
$  4 $ & $\Big(\hfill\frac{1+t}{2},$ & $ \frac{1-t}{2},$ & $-\frac{t}{\sqrt{2}}\Big)$ \\[0.1cm]

$  5 $ & $\Big(-\frac{1+t}{2},     $ & $ \frac{1-t}{2},$ & $-\frac{t}{\sqrt{2}}\Big)$ &
$  6 $ & $\Big(-\frac{1-t}{2},     $ & $ \frac{1+t}{2},$ & $ \frac{t}{\sqrt{2}}\Big)$ \\[0.1cm]

$  7 $ & $\Big(-\frac{1-t}{2},     $ & $-\frac{1+t}{2},$ & $ \frac{t}{\sqrt{2}}\Big)$ &
$  8 $ & $\Big(-\frac{1+t}{2},     $ & $-\frac{1-t}{2},$ & $-\frac{t}{\sqrt{2}}\Big)$ \\[0.1cm]

$  9 $ & $\Big(\hfill\frac{1-t}{2},$ & $-\frac{1+t}{2},$ & $ \frac{t}{\sqrt{2}}\Big)$ &
$ 10 $ & $\Big(\hfill\frac{1+t}{2},$ & $-\frac{1-t}{2},$ & $-\frac{t}{\sqrt{2}}\Big)$ \\[0.1cm]

$ 11 $ & $\Big(\hfill            0,$ & $ t,            $ & $ \frac{1}{\sqrt{2}}\Big)$ &
$ 12 $ & $\Big(\hfill            0,$ & $-t,            $ & $ \frac{1}{\sqrt{2}}\Big)$ \\[0.1cm]
\hline
\end{tabular}
\renewcommand{\arraystretch}{1}
\end{center}
\label{tab1}
\end{table}

As it is shown in Fig.~1, the TTM reduces to the tetrahedral model in the limit $t = 1$, 
apart from the multiplicity of 3 for each tetrahedron vertices. One can check the equivalence 
$\mathbf{s}^{[ 1 ]}_{~}( 1 ) = \mathbf{s}^{[ 4 ]}_{~}( 1 ) = \mathbf{s}^{[ 10 ]}_{~}( 1 )$ 
from Table~I, and the same for the groups $k = 2, 5, 8$, $k = 3, 6, 11$, and $k = 7, 9, 12$. 
The tetrahedral model is essentially equivalent to the 4-state Potts model~\cite{wu}.
In another limit $t = 0$, the TTM reduces to the octahedral model; 
in this case $k = 1, 2$, $k = 3, 4$, $k = 5, 6$, $k = 7, 8$, $k = 9, 10$, and
$k = 11, 12$ are the 6 direction of the octahedron vertices.

In order to obtain the phase diagram with respect to the parameter $t$ and 
the temperature $T$, we calculate the free energy of the TTM.
Let us consider the finite size system of the size $L$ by $L$. The partition function 
\begin{equation}
Z_t^{~}( T; L ) = \sum\limits_{\{\mathbf{S}( t )\}}^{~}
\exp\left( -\frac{H_t}{k_{\rm B}^{~}T} \right) 
\end{equation}
is the configuration sum of Boltzmann factor taken over all the spins denoted 
by $\{\mathbf{S}(t)\}$. Here, $k_{\rm B}^{~}$ is Boltzmann constant, and we use
the dimensionless units by setting $k_{\rm B}^{~}=J=1$. Once $Z_t^{~}( T; L )$
is obtained for a series of system size 
$L$, we can estimate the free energy per site
\begin{equation}
f_t^{~}( T ) = \lim_{L \to \infty}^{~} -\frac{1}{L^2_{~}} \,\, k_{\rm B}T \, \ln Z_t^{~}( T; L )
\end{equation}
in the thermodynamic limit. 

As a numerical tool to obtain $Z_t^{~}( T; L )$, we use the CTMRG method, 
which was developed from Baxter's corner transfer matrix (CTM) formalism~\cite{Baxter}. 
The method enables to calculate the partition function in the form
\begin{equation}
Z_t^{~}( T; L ) =  {\rm Tr}\ C_{~}^4 \, ,
\end{equation}
where $C$ is the CTM, which corresponds to a quadrant
of the finite system~\cite{ctmrg1,ctmrg2}. 
It is convenient to define the normalized density matrix 
\begin{equation}
\rho( T; L ) = 
\frac{ C^4_{~} }{ {\rm Tr} \ C_{~}^4 } = 
\frac{ C^4_{~} }{ Z_t^{~} } \, , 
\label{rho}
\end{equation}
and the mean value of a local operator $O$ at the center of the system 
is given by $\langle O\rangle = {\rm Tr} (O \rho)$. 
In the CTMRG calculations, we keep $m = 300$ representative states at 
most.  Further details of the free energy analysis by CTMRG can be found in Ref.~\onlinecite{Axelrod}. 

A 2D classical system is related to a 1D quantum systems via so called the quantum-
classical correspondence, which is justified via the path integral formulation;~\cite{fradkin} 
on the discrete lattice, the Suzuki-Trotter decomposition provides an explicit 
mapping~\cite{Trotter,Suzuki1,Suzuki2}. This correspondence enables us to introduce the 
notions of quantum information, such as the concurrence~\cite{osterloh,osborne} and the 
entanglement entropy~\cite{osborne,vidal,franchini} to 2D classical systems.
Let us regard the horizontal direction of our 2D classical lattice model as the space direction,
and vertical direction as the imaginary time one. The lower-half lattice is then identified with
the past, and the upper half is the future. In the CTM formalism, both of these halves are 
represented as the product $C^2_{~}$ of two CTMs, and therefore the density matrix 
$\rho( T; L )$ in Eq.~(6) corresponds to square geometry on the 2D lattice, where there is 
a cut from the center of the system toward either left or right boundary with open boundary
condition. The detail of this correspondence is reported by Tagliacozzo {\it et al.}~\cite{tagliacozzo}.

\begin{figure}
\includegraphics[width=0.45\textwidth,clip]{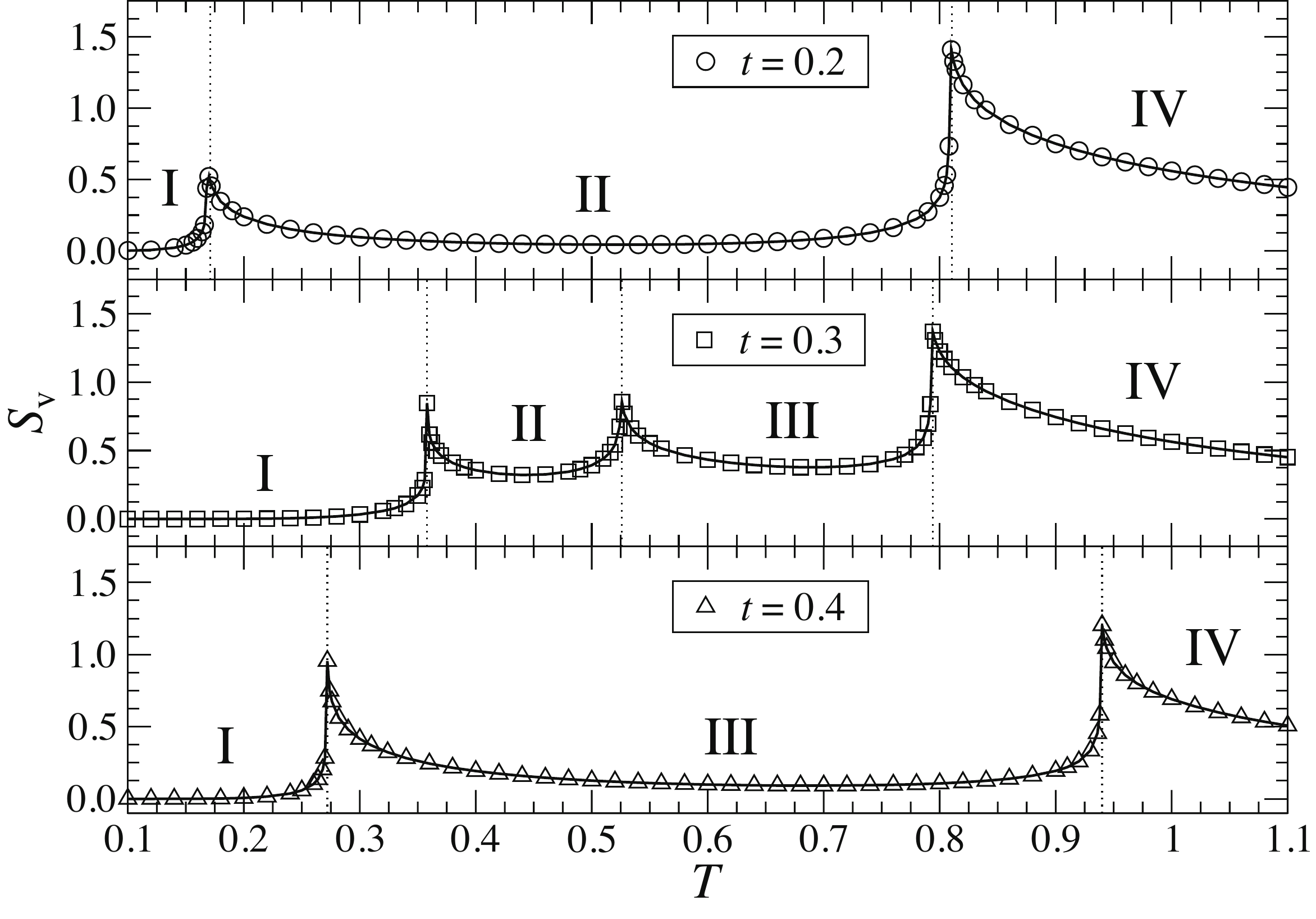}
\caption{
The temperature dependence of the classical analogue of the entanglement entropy
$S_v^{~}$ for $t = 0.2$, $t = 0.3$, and $t = 0.4$. 
The vertical dotted lines denote phase boundaries, and each phase
is labeled by the Roman number.}
\label{fig2}
\end{figure}

Based on the quantum-classical correspondence, the classical analogue of the entanglement 
entropy in the current study is represented as
\begin{equation}
S_v^{~}( T; L ) = - {\rm Tr}\ \rho \ln \rho \,\, \sim \,\, -\sum\limits_{k=1}^{m} \lambda_k^{~} \ln \lambda_k^{~} \, ,
\label{EntEnt}
\end{equation}
where $\lambda_k^{~}$ are the eigenvalues of the density matrix $\rho( T; L )$ in Eq.~(6). 
As a consequence of the conformal invariance at criticality~\cite{CFT}, 
it is known that close to a critical point, the entanglement entropy scales as 
$S_v^{~}( T; \infty ) \sim \frac{c}{6} \, \ln \, \xi$, where $c$ is a central charge and 
$\xi$ is the correlation length in both 1D-quantum and 2D classical 
systems~\cite{Holzhey,korepin,calabrese,ercolessi}; note that we are effectively considering
a system with open boundary condition. Thus $S_v^{~}( T; \infty )$ is divergent at the critical point, 
and can be used for finding the location of phase boundaries~\cite{tagliacozzo}. 
In the case of first-order phase transition, $S_v^{~}( T; \infty )$ is discontinuous at the transition point. 
As examples, we show $S_v^{~}( T; L )$ when $L = 1000$ for the cases $t = 0.2$, $0.3$, and $0.4$
with respect to temperature $T$ in Fig.~\ref{fig2}. It should be noted that there is no need to observe
thermodynamic functions and order parameters for the determination of the phase boundary.

%

%%%%%
\section{Phase Diagram}
%%%%%

\begin{figure}
\includegraphics[width=0.45\textwidth,clip]{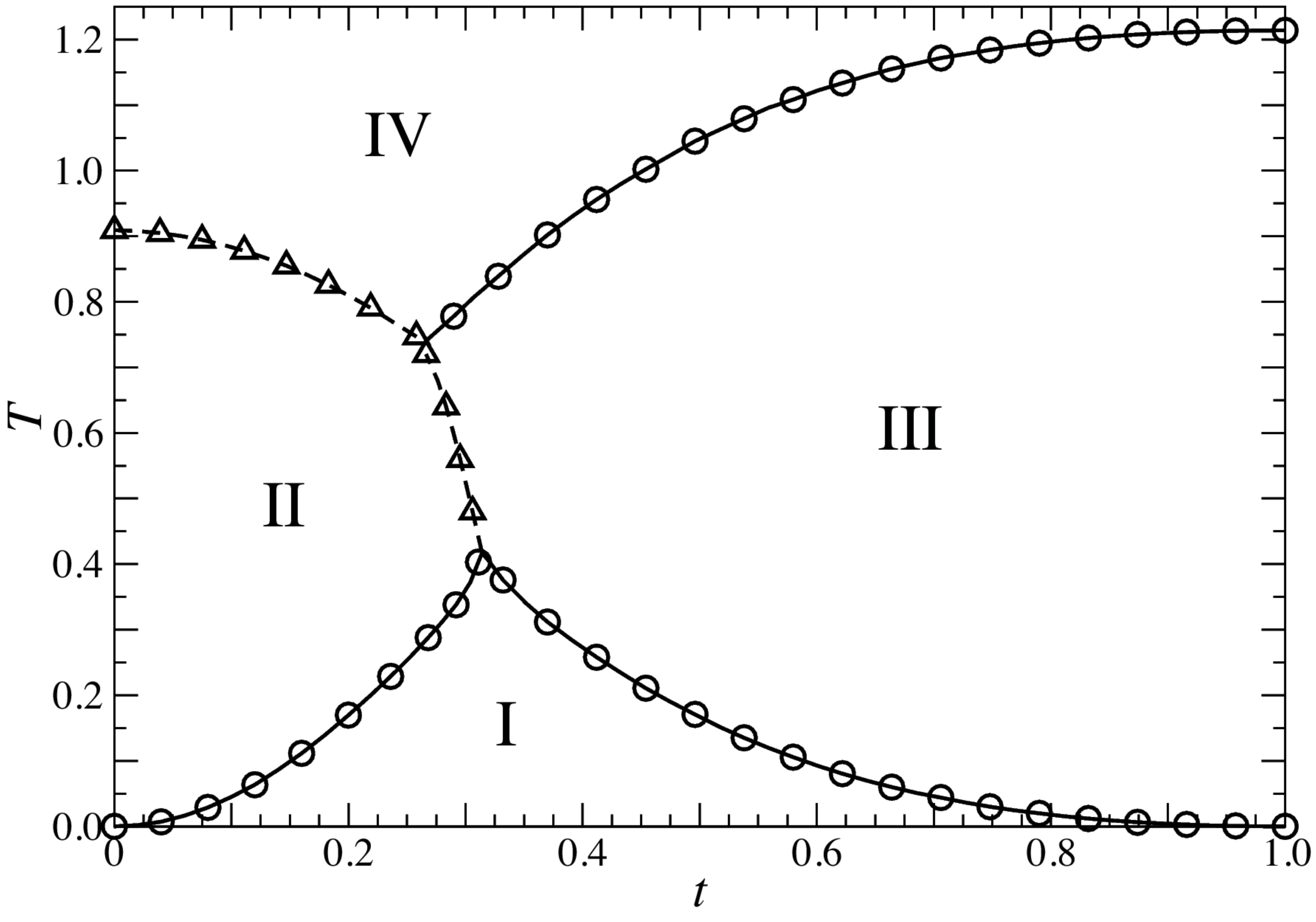}
\caption{
Phase diagram of the TTM with respect to the parameter $t$ and the temperature $T$. 
The circles denote the 2$^{\rm nd}$ order phase transition. The phase boundaries shown by triangles
are identified as the 1$^{\rm st}$ order ones.
}
\label{fig3}
\end{figure}

Figure~\ref{fig3} shows the phase diagram of the TTM determined from the singular or 
discontinuous behavior in $S_v^{~}$ as shown in Fig.~\ref{fig2}.
There are 4 phases, which are labeled from I to IV in the diagram. 
In the low-temperature side, there is a ferromagnetic phase I, where
the symmetry is totally broken. The intermediate phase II in the octahedral side has the
$Z_2^{~}$ symmetry, and if the directions $k = 1$ and $k = 2$ according to Table~I are 
spontaneously chosen, these two directions appear equally.
The intermediate phase III in the tetrahedral side has the $D_3^{~}$ symmetry, 
and if the directions $k = 1$, $k = 4$, and $k = 10$ are spontaneously chosen, these
three directions appear equally. 
The phase IV in the high temperature side is completely disordered. 
The phase boundaries shown by the circles are of the second-order phase transition, and
those shown by the triangles are identified as first-order ones. We observe the detail
of each phase boundary in the following.

\begin{figure}
\includegraphics[width=0.45\textwidth,clip]{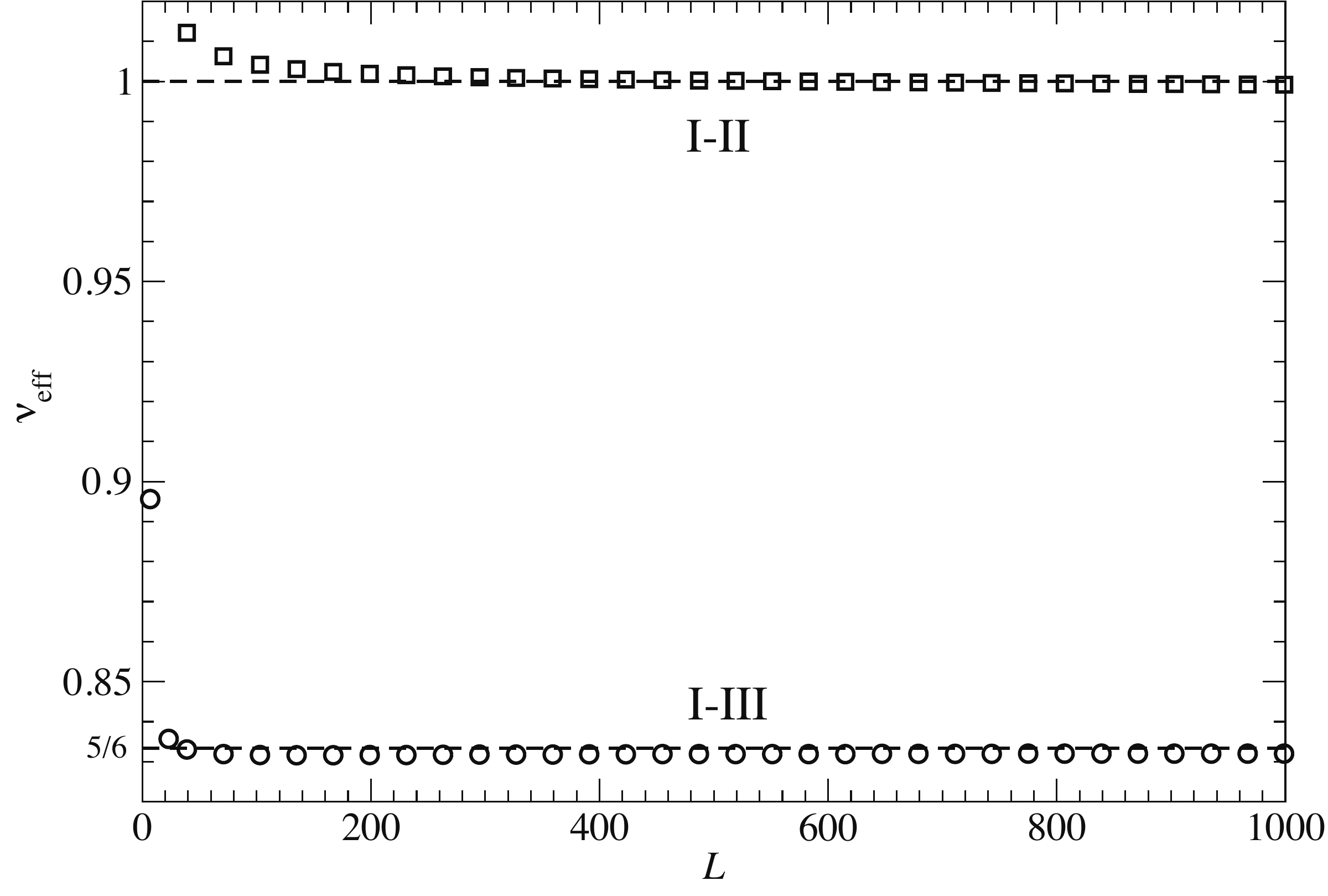}
\caption{The effective  exponent $\nu_{\rm eff}^{~}( L )$ in Eq.~(10) 
calculated at the I--II phase boundary when $t=0.1$ (squares), and 
that at the I--III boundary when $t=0.35$ (circles). }
\label{fig4}
\end{figure}

Let us observe the phase boundary between the phases I and II, 
and also the boundary between the phases I and III.
When the transition is of the second order, its universality can be determined 
by means of finite-size corrections in thermodynamic functions~\cite{fss,tomotoshi}. 
We consider the internal energy per site
\begin{equation}
u_t^{~}( T; L ) = \frac{T^2_{~}}{L^2_{~}} \, \frac{\partial}{\partial T} 
\left\{ k_{\rm B}^{~} \ln Z_t^{~}( T; L ) \right\}
\label{e_n_g}
\end{equation}
as an example. At the critical temperature $T_{\rm c}^{~}$, the internal energy per site satisfies 
\begin{equation}
u_t^{~}( T_{\rm c}^{~}; L ) - u_t^{~}( T_{\rm c}^{~}; \infty ) \, \equiv \, 
\Delta u_t^{~}( L ) \, \propto \, L^{1/\nu - 2}_{~} \, ,
\label{nu}
\end{equation}
where $\nu$ is the scaling exponent for the correlation length.
One can obtain $\nu$  observing the $L$-dependence of the effective  value
\begin{equation}  
\nu_{\rm eff}(L) = \left[ 2 + \frac{\partial\ln\Delta u_t^{~}( L )}{\partial\ln L} \right]^{-1}_{~} \, ,
\end{equation}
which is shown in Fig.~\ref{fig4}. The results agree with the Ising universality with $\nu = 1$ for
the I--II phase boundary ($t = 0.1$), and the 3-state Potts universality with $\nu = 5/6$ for the 
I--III boundary ($t = 0.35$).

\begin{figure}
\includegraphics[width=0.45\textwidth,clip]{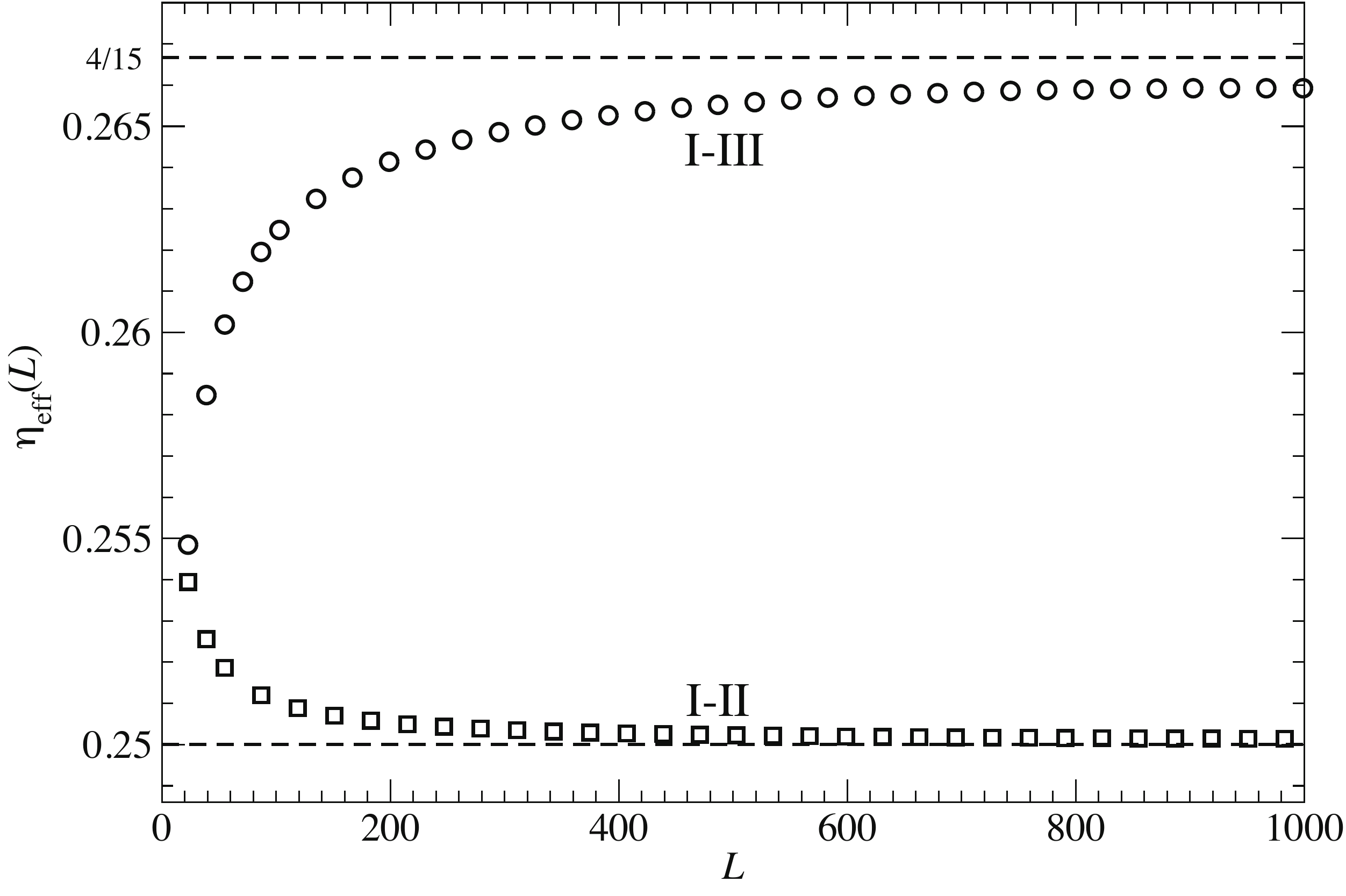}
\caption{The effective exponent $\eta_{\rm eff}^{~}( L )$ in Eq.~(12) calculated at the 
I--II phase boundary when $t = 0.1$ (squares), and that at the I--III boundary when $t=0.35$ (circles). }
\label{fig5}
\end{figure}

In order to obtain another scaling exponent, we observe an appropriate order parameter
$O_t^{~}( T; L )$, for which the finite-size correction satisfies the relation
\begin{equation}
O_t^{~}( T_{\rm c}^{~}; L )
\, \equiv \, \Delta O_t^{~}( L ) \, \propto \, L^{-\eta/2}_{~}
\label{eta}
\end{equation}
noticing that $O_t^{~}( T_{\rm c}^{~}; \infty )$ is zero. In the same manner as we have considered
in Eq.~(10), we can obtain $\eta$ from the effective value
\begin{equation}  
\eta_{\rm eff}^{~}(L) = - 2 \, \frac{\partial\ln \Delta O_t^{~}( L )}{\partial\ln L} \, .
\end{equation}
Inside the phase II we choose the order parameter
\begin{equation}
\begin{split}
O_t^{~}( L ) = & p_1+p_2+p_3+p_4-2p_5-2p_6+\\
             & p_7+p_8-2p_9-2p_{10}+p_{11}+p_{12}\, .
\end{split}
\end{equation}
where $p_k^{~}$ are the probability of the spin at the center of the system to point to
$k$-th direction listed in Table~I. Inside the phase III we choose
\begin{equation}
\begin{split}
O_t^{~}( L ) = & p_1-p_2+p_3-p_4+p_5-p_6+\\
             & p_7-p_8+p_9-p_{10}+p_{11}-p_{12}\, .
\end{split}
\end {equation}
Figure~\ref{fig5} shows the system size dependence of $\eta_{\rm eff}( L )$. 
The behavior at the I--II phase boundary ($t = 0.1$) agrees with Ising 
universality class with $\eta = 1 / 4$. At the I--III boundary, the convergence
with respect to $L$ is rather slow, but $\eta_{\rm eff}( L )$ certainly approaches 
to the value $\eta = 4 / 15$ of the 3-state Potts universality class.

\begin{figure}
\includegraphics[width=0.45\textwidth,clip]{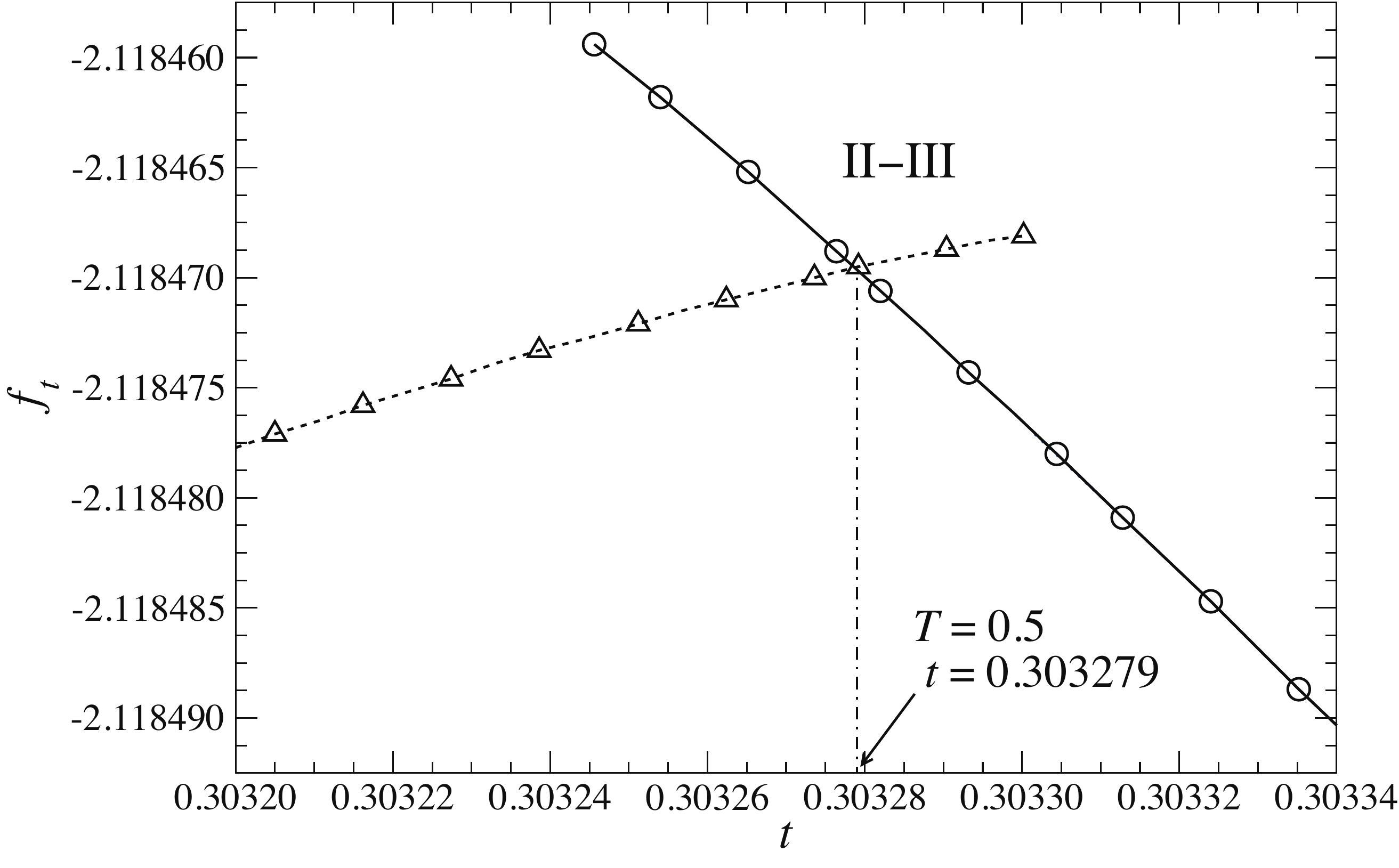}
\caption{
The free energy per site $f_t^{~}( T )$ with respect to the parameter $t$
at fixed temperature $T=0.5$ under the fixed (triangles) and the free (circles)
boundary conditions.
}
\label{fig6}
\end{figure}

We next observe the II--III phase boundary in the intermediate temperature region. 
Figure~\ref{fig6} shows the crossing behavior in the free energy per site $f_t^{~}( T )$ at $T=0.5$
with respect to $t$, where the crossing point is $t = 0.303279$. 
We have chosen both fixed and free boundary conditions to weakly favor one of the two phases.
Since the $Z_2^{~}$ symmetry in the phase II and $D_3^{~}$ symmetry in the phase
III are not a sub-group with each other, a direct second-order phase transition between 
these phases is prohibited. It should be noted that the II--III boundary is not vertical in 
Fig.~3.

\begin{figure}
\includegraphics[width=0.45\textwidth,clip]{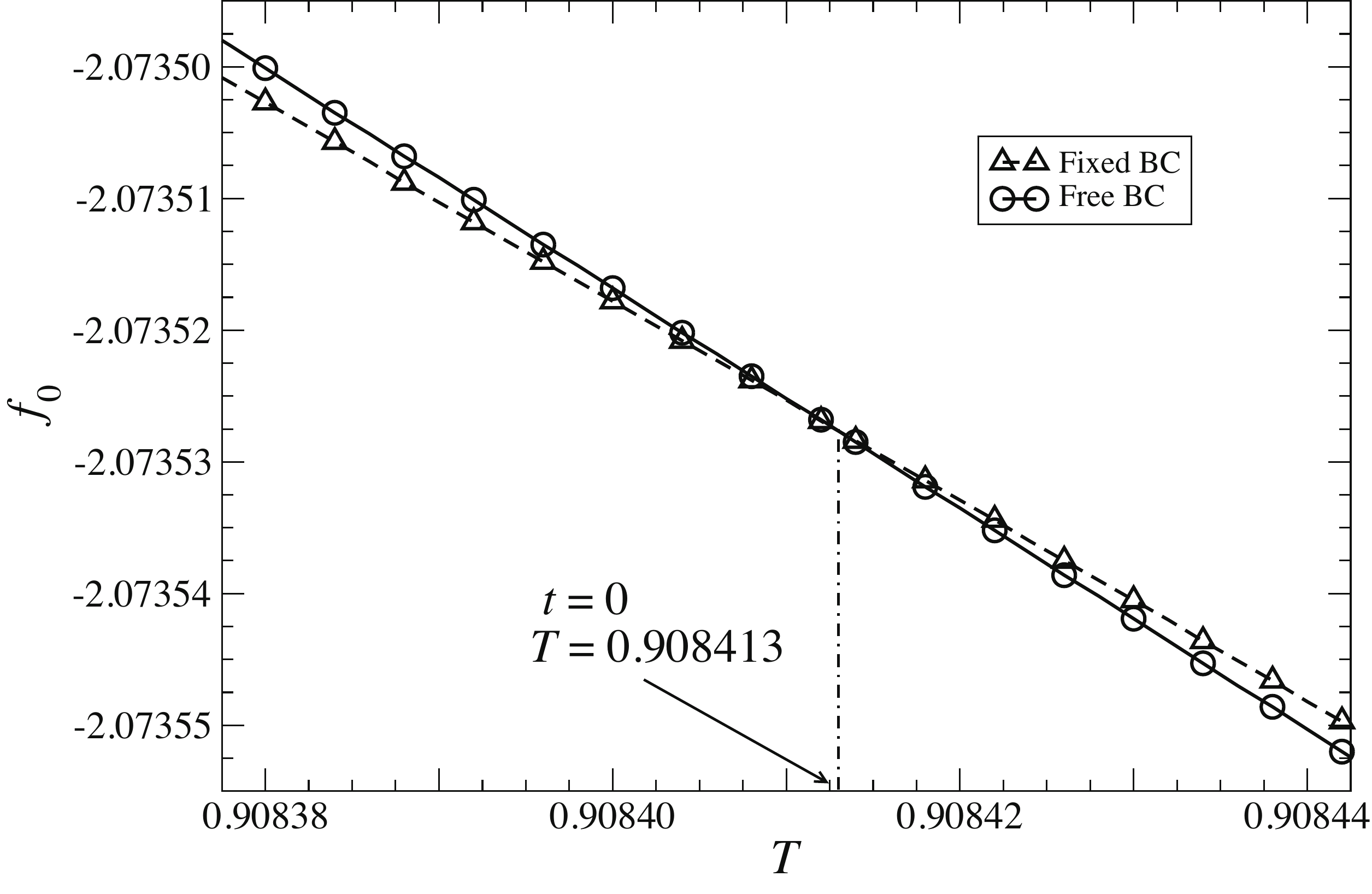}
\caption{
Temperature dependence of the free energy per site $f_{t}^{~}( T )$ in Eq.~(4) 
when $t = 0$, where the TTM coincides with the octahedral model. 
The triangles and the circles correspond to $f_{t}^{~}( T )$ calculated under fixed 
and free boundary condition, respectively. }
\label{fig7}
\end{figure}

\begin{figure}
\includegraphics[width=0.45\textwidth,clip]{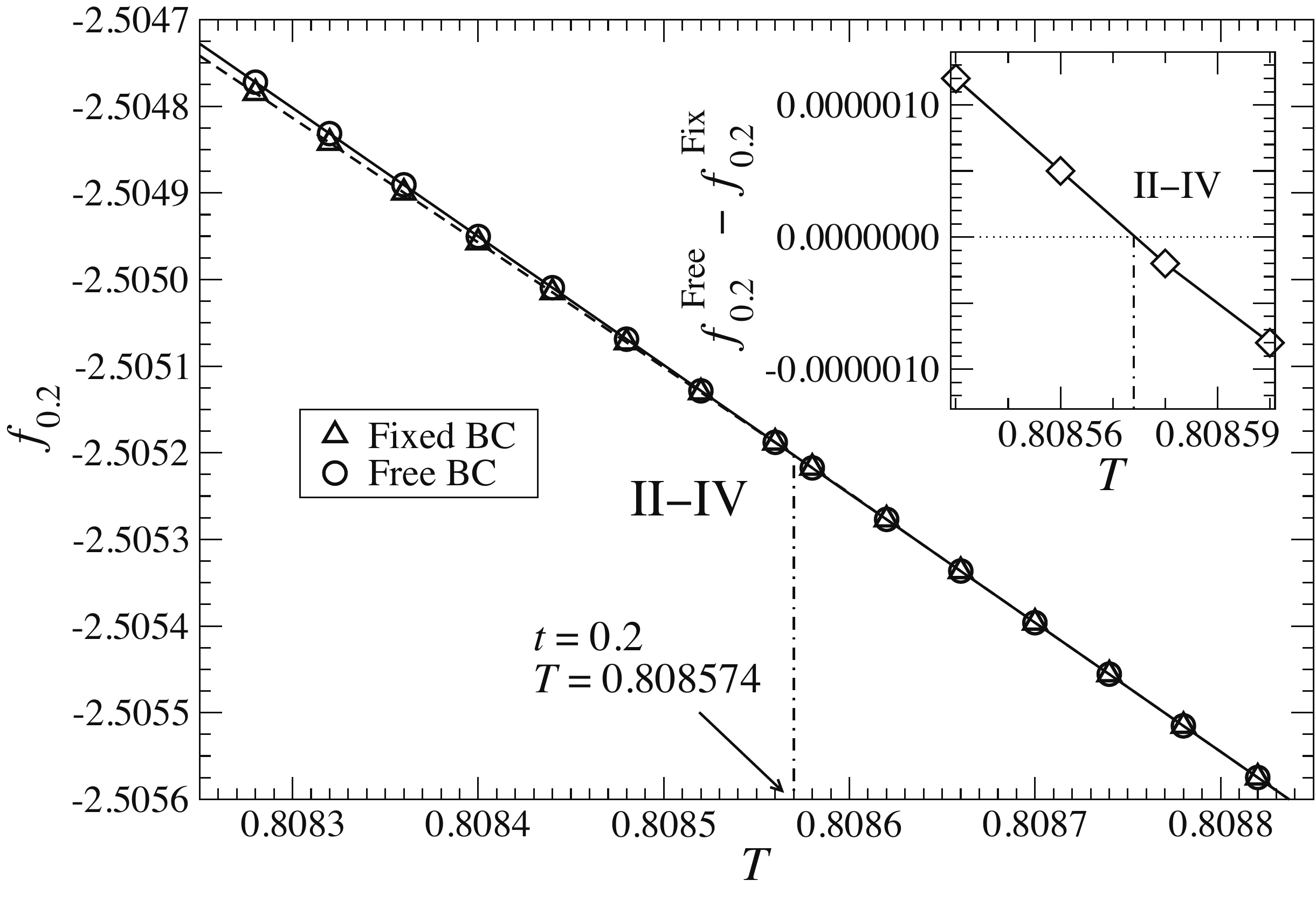}
\caption{The free energy per site $f_{t=0.2}^{~}( T )$ around the II--IV phase boundary 
calculated for both fixed and free boundary conditions; the inset show their difference.
}
\label{fig8}
\end{figure}

Figure~\ref{fig7} shows the calculated free energy per site $f_{t=0}^{~}( T )$ 
at the octahedral limit $t = 0$, which are calculated under the fixed and
the free boundary conditions. Within the shown temperature region, the lower 
plots are thermodynamically stable, and the upper ones are quasi-stable. 
These two lines crosses at $T_{0}^{~} = 0.908413$. 
This crossing behavior in $f_{t=0}^{~}( T )$ shows that the transition is of the first-order.
At the transition temperature $T_0^{~}$, the internal energy per site
$u_t^{~}( T; L )$ in Eq.~(8) is discontinuous in the thermodynamic limit. 
From the jump in $u_t^{~}( T; L )$ when $L$ is sufficiently large, 
we obtain the latent heat $Q = 0.073$.
Figure~\ref{fig8} shows $f_{t=0}^{~}( T )$ when $t = 0.2$. Again we observe
the crossing behavior in $f_t^{~}( T )$, where the lines crosses at $T_{0}^{~} = 0.808574$. 
The latent heat is estimated as $Q = 0.028$. These results support the
presence of a weak first-order phase transition along the II--IV phase boundary.


Since the TTM coincides with the tetrahedral model in the limit $t = 1$, 
the critical temperature $T_{\rm c}^{~}$ in this limit can be calculated exactly as 
$T_{\rm c}^{~} = 4J / (3 \ln 3) \approx 1.21365 \, J$, and the transition belongs
to the 4-state Potts universality class. In this case, numerical confirmation is
not straight forward, because of the nature of BKT
transition~\cite{B1, B2, KT}. Along the III--IV phase boundary, we observed a very slow
convergence in free energy with respect to the system size $L$,
which suggests the presence of BKT transition in the whole part of the III--IV boundary.
We leave the confirmation of this conjecture for a future study.

%%%%%
\section{Discussion and Conclusions}
%%%%%

We have investigated the phase diagram and the thermodynamic properties of the 
truncated tetrahedral model by means of the CTMRG method. It is shown that the
classical analogue of the entanglement entropy, which can be calculated from the 
eigenvalue spectrum of the corner transfer matrix, is efficient for the detection 
of the phase boundaries. Since the free energy per site is directly obtained by 
the CTMRG method, first-order phase transition can be directly detected as a
crossing of the value.

As a result of the numerical calculation, four phases are detected. There is the 
ferromagnetic ``phase I'' in the low temperature side. In the intermediate temperature 
region, there are the ``phase II'' with $Z_2^{~}$ symmetry and the ``phase III'' with 
$D_3^{~}$ one;  the boundary between these intermediate phases is of the first order. 
The phase transition between the completely disordered high-temperature phase, the 
``phase IV'', to the phase II is of the first order, where the calculated latent heat is very 
small. Thus in both octahedral and tetrahedral limits, the effect of the truncation is 
perturbative in the sense that phase II and III occupy finite area in the phase diagram, 
and that the phase I and IV do not touch directly. The BKT transition between the phase
III and IV could be further analyzed by means of modern finite size scaling method by
Hsieh et al.~\cite{sandvik}

The presence of the weak first-order transition on the II-IV phase boundary including 
the octahedral limit $t \rightarrow 0$ draws an attention to revisit both the icosahedral 
and the dodecahedral models, which have larger local degrees of freedom than the 
truncated tetrahedron model we have considered. It should be noted that the two-dimensional 
$q$-state Potts model show first-order phase transition when $q \ge 5$; similar first-order
nature could be expected also in polyhedron models, when the site degrees of freedom is
relatively large. To perform the numerical CTMRG calculation in a stable manner under 
icosahedral or dodecahedral symmetry is a kind of computational challenge,
since the requirements on the computational memory are huge compared with the
currently available computational resources.

\begin{acknowledgments}
This work was supported by the projects QETWORK APVV-14-0878 and
VEGA-2/0130/15. T.~N. and A.~G. acknowledge the support of Grant-in-Aid for
Scientific Research. R.~K. acknowledges the support of Japan Society for Promotion
of Science P12815.
\end{acknowledgments} 

\bibliography{Krcmar_rev.bib}

\end{document}